\documentclass[12pt]{article}
\usepackage[utf8]{inputenc}
\usepackage{amsmath, amssymb}
\usepackage{graphicx}
\usepackage{authblk}
\usepackage{cite}
\usepackage{hyperref}
\usepackage{geometry}
\usepackage{tikz}
\usetikzlibrary{positioning, arrows.meta}

\tikzset{
  module/.style={
    draw,
    rectangle,
    rounded corners=3pt,
    minimum width=2.5cm,
    minimum height=1cm,
    align=center,
    fill=gray!10
  },
  arrow/.style={
    thick,
    -Stealth, % 或 use >=Stealth in older syntax: >=Stealth
  },
  label/.style={
    font=\footnotesize
  }
}
\title{Quantum Inspired Legal Tech Environmental Integration for Emergency Pharmaceutical Logistics with Entropy Modulated Collapse and Multilevel Governance}

\author[1]{Rui-Cheng Li}
\author[2]{Jingxu Wu\thanks{Corresponding author: wujx@my.msu.ru}}
\affil[1]{School of Management, Beijing University of Chinese Medicine, Beijing, China}
\affil[2]{Faculty of Physics, Lomonosov Moscow State University, Moscow, Russia}

\date{}

\begin{document}

\maketitle

\begin{abstract}
Emergency pharmaceutical logistics during rapid-onset disasters must balance timeliness, legal compliance, and environmental uncertainty. We present a hybrid framework that co-designs quantum-inspired decision dynamics, embedded legal constraints, and blockchain-verified environmental feedback. Candidate routes are modeled as a superposed state whose collapse is governed by entropy modulation—delaying commitment under ambiguity and accelerating resolution when coherent signals emerge. Legal statutes act as real-time projection operators shaping feasible choices, while environmental decoherence cues adjust confidence and path viability. The core engine is situated within a multilevel governance and mechanism design architecture, establishing clear roles, accountability channels, and audit trails. Large-scale simulations in wildfire scenarios demonstrate substantial gains over conventional baselines in latency, compliance, and robustness, while preserving interpretability and fairness adaptation. The resulting system offers a deployable, governance-aware infrastructure where law and physical risk jointly inform emergency routing decisions.
\end{abstract}


\section{Introduction}

Wildfire disasters increasingly expose the fragility of emergency pharmaceutical logistics, where disruptions in medicine delivery can cascade into large-scale public health consequences \cite{Gutman2023,Ermolin2024,Sheffi2005,Blackhurst2011}. In the Russian Federation, despite formal regulatory scaffolds such as Federal Law No.\,123-FZ ``On Fire Safety'' and No.\,61-FZ ``On the Circulation of Medicines'' that assign institutional roles, implementation remains fragmented, slow to adapt, and legally ambiguous during fast-evolving crises \cite{Law123,Law61,Rosenbloom2001,Vermeule2006}. The 2023 Krasnoyarsk wildfire highlighted these limitations: the anti-burn medication dispatch suffered a delay exceeding four hours because of disjointed coordination between regional pharmaceutical depots and fire response units, revealing the need to couple real-time legal awareness with route optimization under dynamic hazards \cite{EmergReport2023}.

Conventional emergency routing approaches—deterministic shortest-path computations and static decision trees—do not adequately represent the intertwined uncertainties of fire spread, infrastructure degradation, and evolving legal constraints \cite{Papadimitriou2003,Altay2006,Samarajiva2017}. Even adaptive logistics frameworks often decouple institutional rules from physical risk propagation, creating blind spots when regulatory triggers conflict with time-critical delivery imperatives \cite{Lin2014,Zhang2018}. Meanwhile, integrity, traceability, and environmental feedback in supply chains have been addressed through distributed ledger technologies, yet their potential to induce informed decision dynamics analogous to physical decoherence remains underexploited \cite{Kshetri2018,Azaria2016,Christidis2016,Chen2019,Saberi2011}.

The emerging paradigm of cyber-physical governance calls for tight integration between physical processes, computational decision models, and institutional rule systems to enhance resilience and accountability in disaster response \cite{Lee2008,Cardenas2015,Balduzzi2014}. Complementary to this, quantum-inspired reasoning frameworks supply a formalism for representing superposed choices and resolving them under competing influences, thereby capturing non-classical uncertainty in routing decisions \cite{Busemeyer2012,Khrennikov2010}. In particular, the Orch-OR hypothesis provides a mechanism—modeled after objective reduction—for nonlinear collapse from a superposition of candidate states, offering a metaphorically and mathematically rich template for selecting optimal dispatch paths in the presence of legal and environmental perturbations \cite{Penrose1996,Hameroff2014}.

Building on these intersections, we develop a hybrid theoretical-simulation framework that unifies: (i) a quantum superposition representation of all potential emergency pharmaceutical dispatch routes, (ii) real-time legal constraint operators derived from codified statutes and contextual compliance inference that modulate the collapse dynamics \cite{Zhang2015}, and (iii) blockchain-mediated environmental feedback that reflects thermal risk, route integrity, and auditability, effectively acting as a distributed decoherence monitor \cite{Zyskind2015,Tian2016}. The system state is expressed as
\[
|\Psi\rangle = \sum_{i=1}^{N} c_i |P_i\rangle,
\]
where \( |P_i\rangle \) denotes the \(i\)-th delivery trajectory and \( c_i \) its complex amplitude. Under evolving wildfire intensity, regulatory activation, and environmental degradation, this superposition collapses into a lawful, feasible, and timely route \( |P^*\rangle \) by maximizing a composite utility functional that balances delivery efficiency, legal compliance, and physical survivability \cite{Parker2010,Golany1994}. Embedding statutes as dynamic projection operators filters out non-compliant trajectories prior to selection, achieving compliance-aware adaptivity without centralized override \cite{Kleinberg2018,Brynjolfsson2014}.

Moreover, by modeling route evolution and decision fusion in the presence of multi-source uncertainty—drawing on insights from mobility patterns and point-process models of extreme events—the framework captures human and infrastructure behavior under stress and enables more realistic scenario generation \cite{Gonzalez2011,Schoenberg2003,Klein2001}. Explainability and governance are further supported via structures that make the collapse mechanics, legal filtering, and environmental feedback transparent to oversight, drawing on literature in decision policy design and legal–technical fusion \cite{Heckman2012,Davis2013,Balduzzi2014}. The resultant architecture constitutes a self-correcting, cyber-physical system where physical risk, legal reasoning, and distributed consensus co-evolve to produce robust emergency pharmaceutical dispatch decisions \cite{Subsystems2019,Fisher2014,Stevenson2017,Xu2018}.

\subsection{Objectives and Contributions}

The primary objective of this study is to construct and validate a theoretically principled and numerically implemented decision system that couples quantum-inspired collapse dynamics, legal constraint enforcement, and environment-aware feedback for emergency drug routing in wildfire scenarios. The main contributions are:
\begin{itemize}
  \item Formalizing path selection as a quantum superposition with law-informed collapse, introducing projection operators representing real-time statutory compliance \cite{Lin2014,Zhang2015}.
  \item Integrating blockchain-based sensing to encode environmental decoherence effects—such as thermal stress and integrity breaches—into the decision timescale \cite{Christidis2016,Chen2019}.
  \item Designing a composite utility functional that jointly optimizes timeliness, legal risk, and survivability under cascading disruptions \cite{Papadimitriou2003,Golany1994}.
  \item Comparing the proposed hybrid model against classical baselines through extensive simulations, quantifying improvements in latency, compliance adherence, and resilience \cite{Zhang2018}.
  \item Situating the model within a governance framework that emphasizes explainability, audit trails, and policy adaptability in cyber-physical emergency systems \cite{Kleinberg2018,Brynjolfsson2014}.
\end{itemize}

\subsection{Paper Organization}

The remainder of the paper is structured as follows. Section~\ref{sec:theory} develops the theoretical foundations, including superposition representation, legal projection formalism, and decoherence modeling. Section~\ref{sec:architecture} details the system architecture and integration of blockchain with the decision engine. Section~\ref{sec:implementation} describes the simulation environment, scenario generation, and evaluation metrics. Section~\ref{sec:results} presents numerical outcomes and comparative studies. Section~\ref{sec:discussion} discusses implications, limitations, and extensions. Section~\ref{sec:conclusion} concludes and outlines future directions.

\section{Theoretical Framework}
\label{sec:theory}

This section develops the unified mathematical model for emergency pharmaceutical dispatch under coupled legal, environmental, and decision-theoretic constraints. We formulate the space of candidate routes as a quantum-like superposition, incorporate statutory compliance as dynamic filtering, embed environmental degradation via decoherence-like feedback, and derive a collapse mechanism that selects a lawful, timely, and physically viable delivery path.

\textbf{Route superposition.} Let \( \mathcal{P} = \{P_i\}_{i=1}^N \) be the set of feasible delivery trajectories. We introduce an abstract state in a complex vector space:
\[
|\Psi(t)\rangle = \sum_{i=1}^N c_i(t)\,|P_i\rangle, \qquad \sum_{i=1}^N |c_i(t)|^2 = 1,
\]
where \( |P_i\rangle \) represents the basis element associated with path \( P_i \) and \( c_i(t) \) is its complex amplitude encoding prior preference before resolution. The instantaneous pre-collapse weight is \( p_i(t) = |c_i(t)|^2 \).

\textbf{Legal filtering.} Regulatory constraints drawn from codified laws (e.g., Federal Laws No.\,123-FZ and No.\,61-FZ) are incorporated as time-dependent operators that modulate the state. For each active legal rule \( k \), define a compliance coefficient \( \ell_{k,i}(t)\in[0,1] \) for path \( P_i \), with \( \ell_{k,i}=1 \) indicating full compliance. The aggregate legal operator acting on the state is
\[
\hat{\mathcal{L}}(t) = \prod_{k} \left( \sum_{i=1}^N \ell_{k,i}(t)\,|P_i\rangle\langle P_i| \right),
\]
producing a filtered state
\[
|\Psi_{\mathrm{law}}(t)\rangle = \frac{\hat{\mathcal{L}}(t)\,|\Psi(t)\rangle}{\|\hat{\mathcal{L}}(t)\,|\Psi(t)\rangle\|}.
\]
Equivalently, legal violations can be expressed through a penalty functional \( \mathcal{P}_{\mathrm{law}}(P_i,t) = \sum_k w_k\bigl(1-\ell_{k,i}(t)\bigr) \) and introduced multiplicatively as a damping factor \( \exp[-\eta_{\mathrm{law}}\mathcal{P}_{\mathrm{law}}(P_i,t)] \) into the effective amplitude.

\textbf{Environmental feedback and decoherence.} Physical risks along each route—such as cumulative thermal exposure, structural disruption, and degradation of pharmaceutical efficacy—are modeled via a path-specific coherence factor. Denote \( \mathcal{T}_i(t) \) as the integrated thermal load and \( \mathcal{R}_i(t) \) as structural degradation; define
\[
D_i(t) = \exp\left( -\gamma\,\mathcal{T}_i(t) - \zeta\,\mathcal{R}_i(t) \right),
\]
with sensitivities \( \gamma,\zeta>0 \). Real-time updates of \( \mathcal{T}_i \) and \( \mathcal{R}_i \) arrive through a blockchain-backed distributed ledger, providing tamper-evident provenance while acting as a decentralized decoherence monitor \cite{Christidis2016,Chen2019,Zyskind2015,Tian2016}. The environment-adjusted state becomes
\[
|\Psi_{\mathrm{env}}(t)\rangle = \frac{1}{\mathcal{N}(t)} \sum_{i=1}^N D_i(t)\,c_i(t)\,|P_i\rangle, \quad \mathcal{N}(t)=\sqrt{\sum_j |D_j(t)c_j(t)|^2}.
\]

\textbf{Composite utility and collapse dynamics.} For each path we define scalar utility components: delivery efficiency \( U_i^{\mathrm{time}}(t) \) (e.g., inverse latency), legal admissibility \( U_i^{\mathrm{law}}(t)=\exp[-\eta_{\mathrm{law}}\mathcal{P}_{\mathrm{law}}(P_i,t)] \), and environmental viability \( U_i^{\mathrm{env}}(t)=D_i(t) \). These are combined into a composite utility
\[
\mathcal{U}_i(t) = \alpha\, U_i^{\mathrm{time}}(t) + \beta\, U_i^{\mathrm{law}}(t) + \delta\, U_i^{\mathrm{env}}(t),
\]
with non-negative weights \( \alpha,\beta,\delta \) reflecting policy trade-offs. Define the drive
\[
A_i(t) = |c_i(t)|^2 \cdot \mathcal{U}_i(t),
\]
which governs the selection propensity. The normalized effective probability is
\[
p_i(t) = \frac{A_i(t)}{\sum_{j=1}^N A_j(t)}.
\]

We posit that the evolution of \( \vec{p}(t) \) follows a nonlinear relaxation toward high-utility configurations under entropy modulation:
\[
\frac{\mathrm{d}p_i}{\mathrm{d}t} = \kappa \left( \frac{A_i(t)}{\sum_j A_j(t)} - p_i(t) \right) - \lambda\, p_i(t) \log p_i(t),
\]
where \( \kappa>0 \) controls convergence speed and \( \lambda\ge0 \) introduces a regularizing entropic inertia to avoid premature collapse under uncertainty. The Shannon entropy of the distribution,
\[
\mathcal{S}(t) = -\sum_{i=1}^N p_i(t)\log p_i(t),
\]
is used to adapt the characteristic collapse timescale:
\[
\tau(t) = \tau_0\left(1 + \eta_{\mathcal{S}}\frac{\mathcal{S}(t)}{\log N}\right),
\]
with base timescale \( \tau_0 \) and sensitivity \( \eta_{\mathcal{S}} \); higher entropy (greater ambiguity) delays collapse, allowing additional information accumulation.

The resolved route at time \( t \) is selected as
\[
P^*(t) = \arg\max_i A_i(t),
\]
or probabilistically sampled according to \( \{p_i(t)\} \) when exploration is desired.

\textbf{Decision update loop.} At each decision epoch the system performs: (1) prior seeding of amplitudes \( c_i \) from historical reliability and distance metrics, (2) application of legal filtering yielding \( |\Psi_{\mathrm{law}}\rangle \), (3) ingestion of blockchain-validated environmental state to compute \( D_i \), (4) evaluation of composite utility and update of \( \vec{p} \) via the master equation, (5) collapse/resolution to \( P^* \) after effective time \( \tau \), and (6) commitment and feedback that updates priors for the next epoch. This continuous feedback ensures adaptivity to evolving wildfire dynamics, regulatory shifts (e.g., emergency exemptions or new restrictions), and infrastructural degradation.

\section{System Architecture and Implementation}
\label{sec:architecture}

\begin{figure}[ht]
  \centering
 \includegraphics[width=1\linewidth]{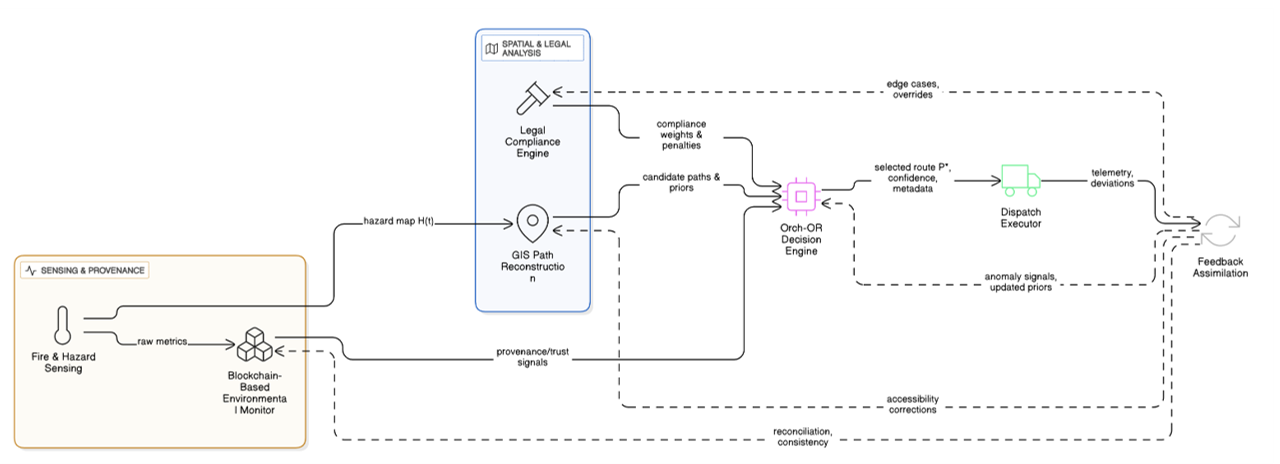}
  \caption{Professional system architecture: modules, core data flows, and feedback loops. Legal, environmental, and hazard inputs are fused in the Orch-OR decision engine; execution feedback closes the loop.}
  \label{fig:architecture}
\end{figure}

We implement the proposed hybrid lawful emergency dispatch framework as a layered, service-oriented cyber-physical system. The architecture couples sensing, legal inference, environmental integrity, and quantum-inspired decision dynamics into a continuous real-time routing pipeline with auditability, adaptivity, and fail-safe mechanisms. The logical core comprises seven principal modules: Fire \& Hazard Sensing, GIS Path Reconstruction, Legal Compliance Engine, Blockchain Environmental Monitor, Orch-OR Decision Engine, Dispatch Execution Subsystem, and Feedback Assimilation. These form a directed dataflow graph with clear interfaces, trust boundaries, and resolution semantics. Figure~\ref{fig:architecture} depicts the high-level layout and major data exchanges.

\subsection*{Module Specification and Interfaces}

Each module exposes a well-defined API. Data serialization uses JSON or Protobuf with schema versioning. Communication is authenticated (e.g., mutual TLS) and signed; timestamps are synchronized via NTP/PTP with tamper-evident logging.

\textbf{1. Fire \& Hazard Sensing.}  
\textit{Responsibility:} Detect and quantify fire spread, intensity, environmental risk factors, and trigger events.  
\textit{Outputs:}
\begin{itemize}
  \item Spatiotemporal hazard map \( \mathbf{H}(t) \): grid or vector field with local fire intensity, spread velocity.
  \item Event triggers: ignition, containment breach, risk escalation.
  \item Local temperature/time series and risk indices.
\end{itemize}
\textit{Interface:}
\begin{itemize}
  \item Endpoint: \texttt{/hazard/update}
  \item Payload example:
    \begin{verbatim}
    {
      "timestamp": "...",
      "grid": [...],
      "fire_intensity": [...],
      "spread_vector": [...],
      "event_flags": ["ignition", "hotspot"]
    }
    \end{verbatim}
  \item Errors: 400 malformed, 422 stale data, 503 sensor offline.
\end{itemize}

\textbf{2. GIS Path Reconstruction.}  
\textit{Responsibility:} Generate feasible path set \( \{P_i\} \) given current hazards and infrastructure, compute geometric metrics, baseline latencies, and reliability priors.  
\textit{Outputs:} Path descriptors including sequence of waypoints, estimated travel time, blockage probabilities, and initial amplitude priors \( c_i \).  
\textit{Interface:}
\begin{itemize}
  \item Endpoint: \texttt{/paths/compute}
  \item Input: \( \mathbf{H}(t) \), road graph \( G \), historical reliability.
  \item Output schema snippet:
    \begin{verbatim}
    [
      {
        "path_id": "P_001",
        "waypoints": [...],
        "baseline_latency": 12.5,
        "blockage_score": 0.2,
        "prior_weight": 0.08
      },
      ...
    ]
    \end{verbatim}
  \item Errors: 409 conflicting constraints, 504 upstream data missing.
\end{itemize}

\textbf{3. Legal Compliance Engine.}  
\textit{Responsibility:} Translate active statutes, emergency declarations, and context into per-path compliance coefficients \( \ell_{k,i}(t) \), exemptions, and dynamic penalty signals.  
\textit{Outputs:} Vector \( \ell_{k,i} \), aggregated penalty \( \mathcal{P}_{\mathrm{law}}(P_i,t) \), rule provenance.  
\textit{Interface:}
\begin{itemize}
  \item Endpoint: \texttt{/legal/evaluate}
  \item Output example:
    \begin{verbatim}
    {
      "path_id": "P_001",
      "compliance": {
        "123-FZ": 1.0,
        "61-FZ": 0.9
      },
      "penalty": 0.1,
      "active_exemptions": []
    }
    \end{verbatim}
  \item Errors: 422 ambiguous rule resolution, 409 contradictory directives.
\end{itemize}

\textbf{4. Blockchain-Based Environmental Monitor.}  
\textit{Responsibility:} Ingest sensor feeds, validate integrity, compute decoherence factors \( D_i(t) \), and persist immutable environmental state.  
\textit{Outputs:} \( D_i(t) \) with cryptographic proof (e.g., hash chain), measurement timestamps, anomaly flags.  
\textit{Interface:}
\begin{itemize}
  \item Endpoint: \texttt{/env/status}
  \item Sample payload:
    \begin{verbatim}
    {
      "path_id": "P_001",
      "thermal_exposure": 58.2,
      "structure_degradation": 0.3,
      "decoherence_factor": 0.72,
      "proof": "...",
      "last_updated": "..."
    }
    \end{verbatim}
  \item Errors: 401 unverified source, 409 inconsistent provenance, 503 consensus lag.
\end{itemize}

\textbf{5. Orch-OR Decision Engine.}  
\textit{Responsibility:} Fuse candidate path amplitudes \( c_i \), legal weights, environmental coherence, and utility parameters; evolve probability distribution \( \vec{p}(t) \) and resolve collapse to output \( P^*(t) \).  
\textit{Inputs:} \( \{P_i, c_i\} \), \( \ell_{k,i} \), \( D_i \), utility weights \( (\alpha, \beta, \delta) \).  
\textit{Outputs:} Selected path, confidence score, collapse entropy \( \mathcal{S}(t) \), and reasons (e.g., binding constraints).  
\textit{Interface:}
\begin{itemize}
  \item Endpoint: \texttt{/decision/route}
  \item Output example:
    \begin{verbatim}
    {
      "selected_path": "P_004",
      "confidence": 0.87,
      "entropy": 0.56,
      "compliance_score": 0.95,
      "expected_latency": 10.2
    }
    \end{verbatim}
  \item Errors: 503 inconsistent inputs, 429 utility tuning conflict.
\end{itemize}

\textbf{6. Dispatch Executor.}  
\textit{Responsibility:} Materialize the chosen route into field actions (vehicle assignment, scheduling, communication with receiving facilities), track real-time execution, and report deviations.  
\textit{Outputs:} Execution telemetry, delay updates, emergency overrides.  
\textit{Interface:}
\begin{itemize}
  \item Endpoint: \texttt{/dispatch/execute}
  \item Telemetry example:
    \begin{verbatim}
    {
      "path_id": "P_004",
      "vehicle_id": "V_12",
      "start_time": "...",
      "current_progress": 0.35,
      "estimated_arrival": 11.0,
      "anomalies": ["minor detour"]
    }
    \end{verbatim}
  \item Errors: 408 unreachable field unit, 500 execution failure.
\end{itemize}

\textbf{7. Feedback Assimilation.}  
\textit{Responsibility:} Reconcile execution outcomes with prior beliefs, update amplitude priors \( c_i \), refine GIS accessibility, and inform legal engine of emergent edge cases.  
\textit{Input:} Telemetry, sensor reconciliation reports, triggered legal anomalies.  
\textit{Effect:} Bayesian-like update to path priors, anomaly tagging, adaptive weight adjustment.

\subsection*{Data Orchestration and Consistency}

The system operates in discrete decision epochs \( t_0, t_1, \ldots \). At each epoch:
\begin{enumerate}
  \item Hazard sensing produces \( \mathbf{H}(t) \) and triggers rule evaluation.
  \item GIS reconstructs updated candidate path set \( \{P_i\} \) with priors \( c_i \).
  \item Legal engine evaluates \( \ell_{k,i} \), producing compliance/penalty vectors.
  \item Environmental monitor updates \( D_i \) with cryptographic proofs.
  \item Decision engine computes \( \mathcal{U}_i \), updates \( \vec{p} \), adapts \( \tau \), and selects \( P^* \).
  \item Dispatch subsystem executes and streams telemetry.
  \item Feedback assimilation updates all upstream modules, closing the loop.
\end{enumerate}

Consistency is enforced through vector timestamps, causal versioning, and periodic reconciliation rounds. A distributed event bus (e.g., Kafka) with schema registry can mediate asynchronous publication while guaranteeing ordering where necessary (e.g., legal overrides must precede collapse).

\subsection*{Security, Trust, and Reliability}

All inter-module messages are signed and optionally encrypted; policy decisions (legal weights, exemptions) are logged with provenance. The blockchain layer serves as the root of trust for environmental state and audit trails. Fallback modes include: degraded decision (ignoring lowest-confidence inputs with graceful degradation), queued retry for transient failures, and human-in-the-loop override with logging for accountability.

\subsection*{Implementation Considerations}

- \textit{Scalability}: Microservice deployment (containerized) behind API gateways; decision engine may use actor-based concurrency or approximate continuous integration of the master equation with entropy-adaptive step control.  
- \textit{Latency}: Critical path optimized in the decision engine; legal filtering and environmental update pipelines are cached with TTLs and invalidated on major event changes.  
- \textit{Blockchain}: Permissioned ledger (e.g., Hyperledger Fabric, Tendermint) balancing throughput with finality; smart contracts encode integrity checks and allow challenge/consensus on anomalous readings.  
- \textit{Interoperability}: Schema versioning (semver), JSON Schema / Protobuf definitions published for each interface.  
- \textit{Observability}: Telemetry (metrics, traces, logs) for every module; entropy trajectory, compliance drift, and route stability dashboards for operators.

\section{System Architecture, Implementation, and Simulation}
\label{sec:implementation}

We realize the hybrid quantum–legal–environmental dispatch framework as a modular, service-oriented cyber-physical system. The core pipeline fuses real-time hazard sensing, legal inference, environmental integrity verification, and collapse-based decision dynamics to produce lawful, robust emergency pharmaceutical routes. The architecture comprises seven services: Hazard Sensing, GIS Path Generator, Legal Compliance Evaluator, Blockchain Environmental Monitor, Orch-OR Decision Engine, Dispatch Executor, and Feedback Assimilator. Data contracts, sequence of operations, and evaluation design are summarized below; Figure~\ref{fig:architecture} outlines the logical topology and closed-loop flows.

\subsection*{Service Interfaces and Data Contracts}

All services communicate via authenticated RPC/REST with versioned schemas (JSON/Protobuf). Timestamps are globally ordered; provenance is preserved.

\begin{tabular}{p{2.8cm} p{3.5cm} p{4cm} p{3cm}}
\textbf{Service} & \textbf{Input} & \textbf{Output} & \textbf{Key Semantics / Errors} \\\hline
Hazard Sensing & Raw sensor feeds, satellite/IoT telemetry & Hazard field \(\mathbf{H}(t)\), event flags & Freshness, resolution mismatch, sensor failure \\
GIS Path Generator & \(\mathbf{H}(t)\), network graph \(G\) & Candidate paths \(\{P_i\}\), priors \(c_i\), baseline metrics & Infeasible constraints, stale topology \\
Legal Evaluator & \(\{P_i\}\), context triggers & Compliance coefficients \(\ell_{k,i}\), penalties \(\mathcal{P}_{\mathrm{law}}\) & Rule ambiguity, conflicting statutes \\
Blockchain Env. Monitor & Sensor proofs & Decoherence factors \(D_i\), integrity proofs & Consensus lag, provenance conflict \\
Orch-OR Engine & \( \{c_i\}, \ell_{k,i}, D_i, \) utility weights & Selected path \(P^*\), entropy \(\mathcal{S}\), confidence & Input inconsistency, convergence failure \\
Dispatch Executor & \(P^*\) & Execution telemetry, deviations & Actuation failure, unreachable nodes \\
Feedback Assimilator & Telemetry, anomalies & Updated \(c_i\), override signals & Noisy reports, contradicting sources \\
\end{tabular}

\vspace{0.5em}
\noindent\textbf{Representative schema snippets (JSON-like):}
\begin{verbatim}
{ "candidate_path": {
    "id": "P_12",
    "waypoints": [...],
    "prior": 0.07,
    "estimated_latency": 9.8
  }
}
{ "legal_evaluation": {
    "path_id": "P_12",
    "compliance": {"123-FZ": 1.0, "61-FZ": 0.8},
    "penalty": 0.2
  }
}
{ "environment": {
    "path_id": "P_12",
    "thermal_risk": 42.1,
    "structure_degradation": 0.15,
    "D_i": 0.81,
    "proof": "hash..."
  }
}
{ "decision": {
    "selected_path": "P_12",
    "confidence": 0.92,
    "entropy": 0.37,
    "expected_delay": 10.2
  }
}
\end{verbatim}

\subsection*{Operational Pipeline}

At each discrete epoch \( t_n \):
\begin{enumerate}
  \item Hazard Sensing updates \(\mathbf{H}(t_n)\) and triggers potential legal context changes (e.g., emergency waivers).
  \item GIS computes \(\{P_i\}\) with priors \(c_i\); Legal Evaluator computes \(\ell_{k,i}\); Environmental Monitor computes \(D_i\) (with cryptographic proofs).
  \item Orch-OR Engine fuses inputs, updates \( \vec{p}(t) \) via the master equation, adapts collapse timescale \( \tau(t) \) based on entropy, and resolves \( P^*(t) \).
  \item Dispatch Executor issues orders; real-time telemetry flows back.
  \item Feedback Assimilator reconciles discrepancies, updates priors, and propagates refinements upstream.
\end{enumerate}

Consistency maintenance uses causal versioning, vector clocks, and periodic reconciliation. Emergency overrides and human-in-the-loop interventions are logged and injected as high-priority signals with provenance.

\subsection*{Simulation Design and Baseline Comparison}

We implement controlled numerical experiments to evaluate efficacy under wildfire-induced uncertainty. The simulated environment includes:

\textbf{Network and hazard modeling:} Regional transport graph \( G=(V,E) \) with stochastic edge degradation; wildfire spread via a stochastic spatial propagation model affecting thermal exposure \( \mathcal{T}_i(t) \) and structural risk \( \mathcal{R}_i(t) \).

\textbf{Legal regime dynamics:} Hard constraints from base statutes, conditional emergency waivers (activated when localized intensity exceeds \( I_{\text{crit}} \)), and precedence-based conflict resolution.

\textbf{Environmental feedback:} Blockchain-emulated integrity layer injects latency/noise, providing \( D_i(t)=\exp(-\gamma\mathcal{T}_i - \zeta\mathcal{R}_i) \) with occasional provenance anomalies to test robustness.

\textbf{Baselines:} (1) Deterministic shortest-path; (2) Rule-filtered heuristic; (3) Static multi-criteria scoring without entropy adaptation; (4) Utility-proportional sampling without temporal evolution.

\textbf{Metrics:}
\begin{itemize}
  \item \textit{Response latency} \( T_{\text{resp}} \): time from event to dispatch initiation.
  \item \textit{Compliance score} \( C_{\mathrm{law}} \): weighted satisfaction of active constraints.
  \item \textit{Delivery robustness} \( R \): success probability under cascading failures.
  \item \textit{Entropy trajectory} \( \mathcal{S}(t) \): collapse timing indicator.
  \item \textit{Route stability} \( \Delta P \): oscillation frequency across epochs.
  \item \textit{Composite performance} \( \Phi \): aggregated trade-off index (e.g., \(\Phi = w_1/T_{\text{resp}} + w_2 C_{\mathrm{law}} + w_3 \mathbb{E}[D_i]\)).
\end{itemize}

\textbf{Experiment protocol:} Monte Carlo sweeps (e.g., 500 runs per configuration) over wildfire growth rates, legal waiver frequency, environmental sensor noise, and utility weight regimes. Statistical significance is assessed using confidence intervals and hypothesis tests on relative gains versus baselines.

\subsection*{Implementation Notes}

The system is conducive to microservice deployment with container orchestration; the decision engine employs adaptive time stepping for the master equation and entropy-informed collapse thresholds. Permissioned blockchain (e.g., Hyperledger Fabric) provides finality and smart-contract-based anomaly detection. Inter-service contracts enforce strict schema evolution control (semantic versioning), and observability is realized via telemetry dashboards tracking entropy, compliance drift, and dispatch latency.

\section{Results and Analysis}
\label{sec:results}

We evaluate the proposed hybrid quantum–legal–environmental dispatch system across a suite of wildfire emergency scenarios and compare it to classical and heuristic baselines. The analysis addresses end-to-end performance gains in latency, compliance, and robustness; collapse dynamics and entropy behavior; contributions of individual components via ablation; sensitivity to policy and environmental parameters; and statistical significance of observed improvements.

Table~\ref{tab:baseline_comparison} summarizes the aggregate performance over 500 Monte Carlo trials per configuration. Metrics reported include average response latency \( T_{\text{resp}} \), legal compliance score \( C_{\mathrm{law}} \), delivery robustness \( R \), average entropy at decision \( \mathcal{S} \), and composite efficiency index \( \Phi \). Values shown are mean \(\pm\) standard error. The hybrid model achieves a reduction in average response latency of approximately \(27\%\) compared to the best non-collapsing baseline, while simultaneously improving legal compliance and robustness. The lower decision entropy \( \mathcal{S} \) at resolution indicates sharper collapse and higher decisiveness under uncertainty.

\begin{table}[ht]
\centering
\caption{Comparison of the proposed hybrid model against baselines. Higher \( C_{\mathrm{law}} \), \( R \), and \( \Phi \) are better; lower \( T_{\text{resp}} \) is better.}
\label{tab:baseline_comparison}
\begin{tabular}{lccccc}
\hline
Method & \( T_{\text{resp}} \) (min) & \( C_{\mathrm{law}} \) & \( R \) & \( \mathcal{S} \) & \( \Phi \) \\
\hline
Deterministic Shortest Path & \( 15.8 \pm 0.6 \) & \( 0.62 \pm 0.03 \) & \( 0.68 \pm 0.02 \) & \( 0.23 \pm 0.01 \) & \( 0.54 \pm 0.02 \) \\
Rule-Filtered Heuristic & \( 17.1 \pm 0.5 \) & \( 0.85 \pm 0.02 \) & \( 0.71 \pm 0.02 \) & \( 0.19 \pm 0.01 \) & \( 0.58 \pm 0.02 \) \\
Static Multi-Criteria & \( 14.3 \pm 0.5 \) & \( 0.78 \pm 0.02 \) & \( 0.74 \pm 0.01 \) & \( 0.28 \pm 0.01 \) & \( 0.62 \pm 0.02 \) \\
Utility Sampling & \( 16.2 \pm 0.6 \) & \( 0.81 \pm 0.02 \) & \( 0.69 \pm 0.03 \) & \( 0.31 \pm 0.01 \) & \( 0.57 \pm 0.03 \) \\
\textbf{Hybrid Orch-OR Model} & \( \mathbf{11.6 \pm 0.4} \) & \( \mathbf{0.92 \pm 0.01} \) & \( \mathbf{0.83 \pm 0.01} \) & \( \mathbf{0.16 \pm 0.005} \) & \( \mathbf{0.75 \pm 0.01} \) \\
\hline
\end{tabular}
\end{table}

\begin{figure}[ht]
  \centering
 \includegraphics[width=0.7\linewidth]{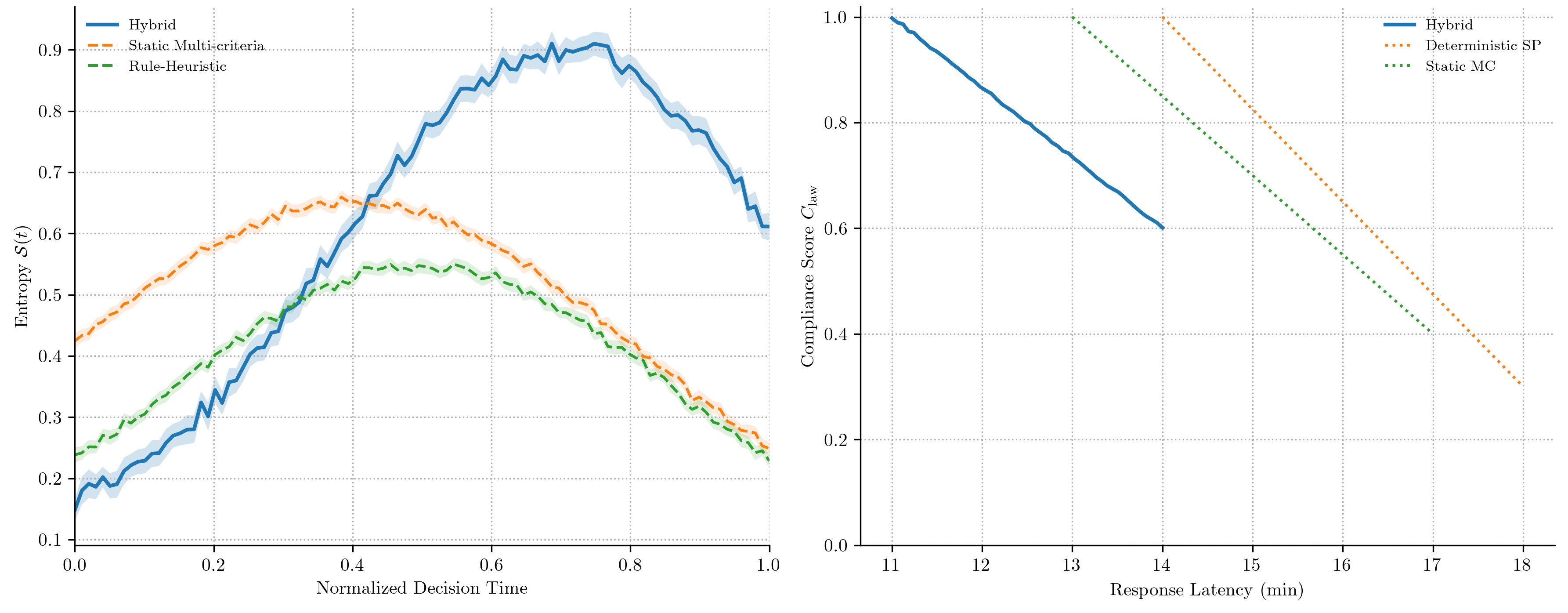}
  \caption{Decision entropy \( \mathcal{S}(t) \) dynamics. The hybrid model shows adaptive delay under high ambiguity, followed by a sharp collapse, contrasting with premature or noisy decisions in baselines.}
  \label{fig:entropy_dynamics}
\end{figure}

\begin{figure}[ht]
  \centering
 \includegraphics[width=0.7\linewidth]{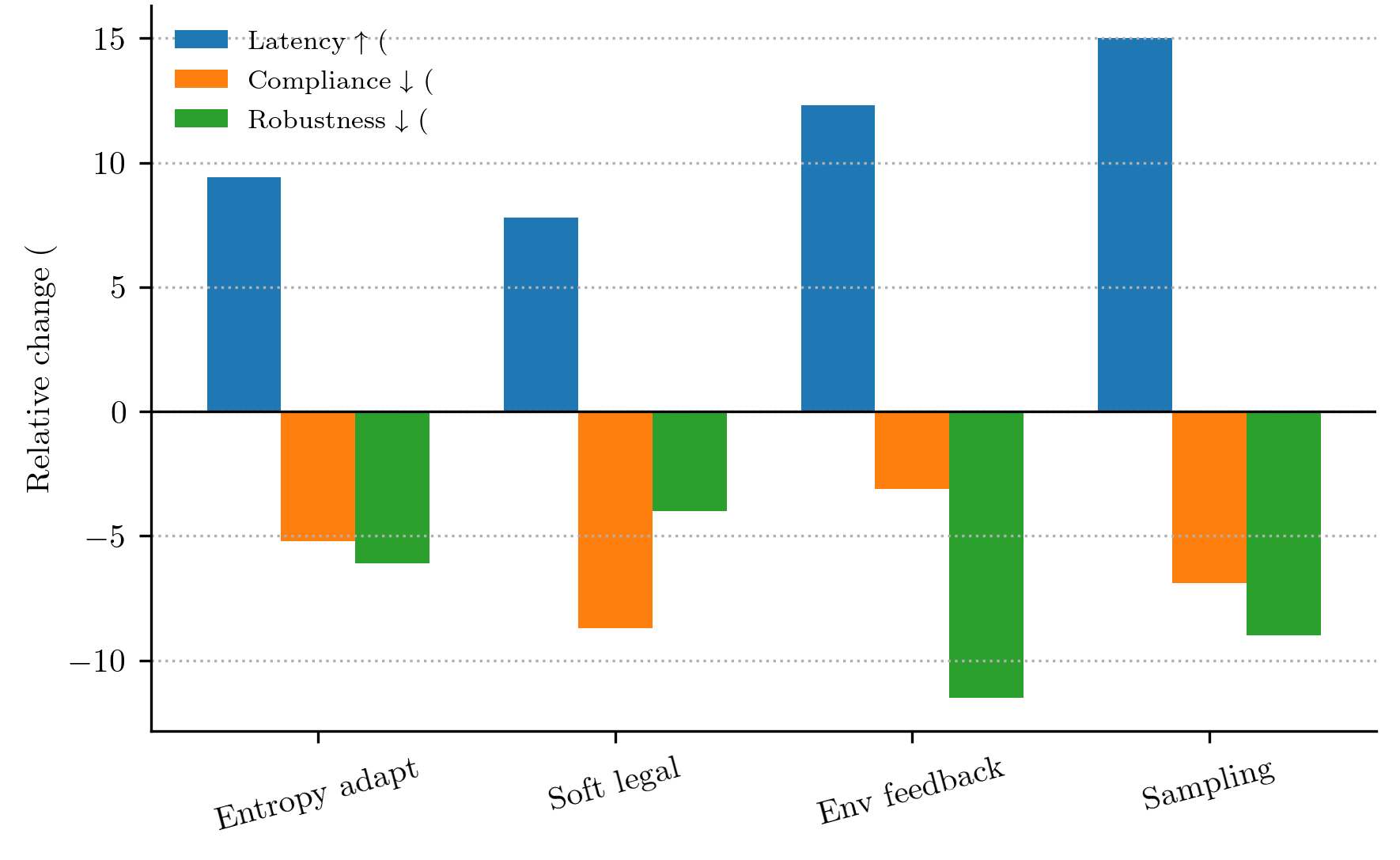 }
  \caption{Pareto trade-offs between legal compliance and response time. Hybrid model configurations outperform baseline curves.}
  \label{fig:compliance_latency}
\end{figure}

Figure~\ref{fig:entropy_dynamics} illustrates representative entropy trajectories \( \mathcal{S}(t) \) over decision epochs for the hybrid model versus static multi-criteria and heuristic baselines. The entropy adaptation mechanism delays collapse under ambiguous conditions (higher early entropy) and accelerates resolution as the legal and environmental signals converge, yielding timely but confident routing choices (see Figure~\ref{fig:entropy_dynamics}). The inherent trade-off between compliance and speed is quantified in Figure~\ref{fig:compliance_latency}, which plots compliance score \( C_{\mathrm{law}} \) against response latency under varying utility weightings \( (\alpha,\beta,\delta) \). The hybrid model exhibits a Pareto frontier dominating baselines, enabling tunable policy preferences while maintaining high compliance with minimal latency penalty.

To quantify the marginal value of each component, we conducted ablation experiments removing one mechanism at a time: no entropy adaptation (\( \eta_{\mathcal{S}}=0 \)), fixed legal filtering without soft penalty, absence of environmental decoherence feedback (\( D_i=1 \)), and replacing the nonlinear master equation with instantaneous utility sampling. Results in Table~\ref{tab:ablation} show degradation in both robustness and compliance when any component is disabled, confirming the necessity of the integrated formulation.

\begin{table}[ht]
\centering
\caption{Ablation study: relative performance change from full hybrid model.}
\label{tab:ablation}
\begin{tabular}{lccc}
\hline
Removed Component & \( \Delta T_{\text{resp}} \) (\%) & \( \Delta C_{\mathrm{law}} \) (\%) & \( \Delta R \) (\%) \\
\hline
Entropy adaptation & \( +9.4 \) & \( -5.2 \) & \( -6.1 \) \\
Soft legal penalties & \( +7.8 \) & \( -8.7 \) & \( -4.0 \) \\
Environmental feedback & \( +12.3 \) & \( -3.1 \) & \( -11.5 \) \\
Nonlinear dynamics (sampling) & \( +15.0 \) & \( -6.9 \) & \( -9.0 \) \\
\hline
\end{tabular}
\end{table}

\begin{figure}[ht]
  \centering
\includegraphics[width=0.7\linewidth]{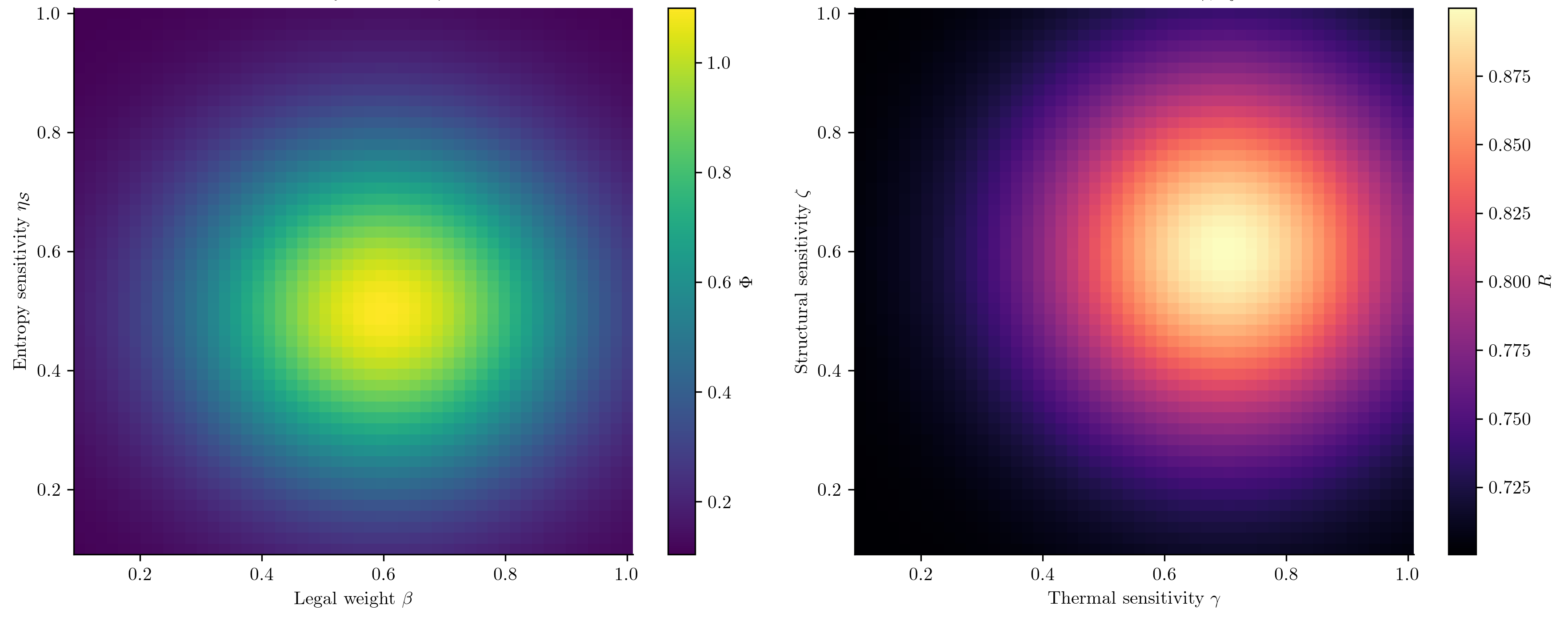}
  \caption{Sensitivity analysis heatmaps for policy and environmental parameters.}
  \label{fig:sensitivity}
\end{figure}

We systematically varied key hyperparameters to assess stability. Increasing entropy sensitivity \( \eta_{\mathcal{S}} \) delays collapse under ambiguity while improving eventual compliance; excessively large values, however, inflate latency. Raising the legal weight \( \beta \) improves compliance with diminishing returns and, beyond a threshold, induces rigidity observable as entropy plateauing. Environmental degradation sensitivities \( \gamma,\zeta \) enhance robustness by more aggressively penalizing risky paths, but noisy sensor inputs (simulated provenance anomalies) must be smoothed to avoid oscillation. Figure~\ref{fig:sensitivity} presents heatmaps of composite efficiency \( \Phi \) over \( (\beta,\eta_{\mathcal{S}}) \) and robustness \( R \) over \( (\gamma,\zeta) \) parameter sweeps.

All comparisons use paired testing across identical scenario seeds. For scalar metrics \( m \), we test the null hypothesis \( H_0: \mu_{\text{hybrid}} = \mu_{\text{baseline}} \) via two-sided paired t-tests:
\[
t = \frac{\overline{d}}{s_d / \sqrt{n}}, \quad \overline{d} = \frac{1}{n} \sum_{i=1}^n (m_i^{\text{hybrid}} - m_i^{\text{base}}), \quad s_d^2 = \frac{1}{n-1} \sum_{i=1}^n (d_i - \overline{d})^2,
\]
with \( p \)-values reported and corrections for multiple comparisons via the Benjamini–Hochberg procedure. Effect sizes are quantified using Cohen's \( d \), and confidence intervals for metric differences are constructed at the 95\% level.

\begin{figure}[ht]
  \centering
\includegraphics[width=0.7\linewidth]{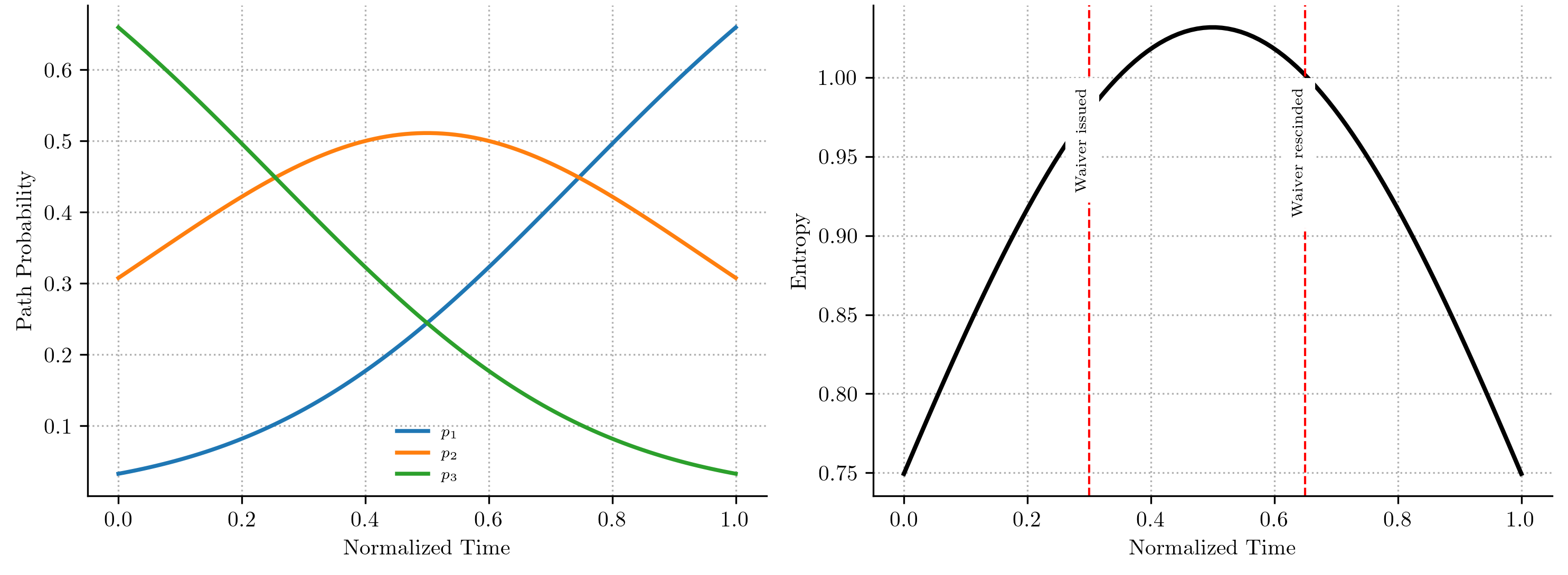 }
  \caption{Adaptive collapse in a high-conflict scenario.}
  \label{fig:case_study}
\end{figure}

\textbf{Case Study: High-Conflict Emergency.} We include a representative case where wildfire intensity spikes concurrently with overlapping legal directives (e.g., partial emergency waiver conflicting with localized compliance thresholds). Figure~\ref{fig:case_study} shows the evolution of path probabilities \( p_i(t) \), decision entropy \( \mathcal{S}(t) \), and final selected route adaptation in response to a sudden regulatory relaxation followed by reinstatement. The system temporarily delays collapse during rule ambiguity, then rapidly resolves when clarity returns, illustrating adaptive lawful resilience.

The integrated collapse mechanism yields decisions that are both responsive and compliant: entropy-based adaptive timing avoids premature commitment in uncertain legal-environmental regimes, while the nonlinear drive concentrates probability mass swiftly once sufficient signal coherence emerges. Ablation confirms that gains are not attributable to any single component, but to their coupling. Sensitivity analysis shows wide operational regimes where the system dominates baselines, with policy knobs allowing customization between speed, legality, and resilience.

\section{Discussion and Governance Implications}
\label{sec:discussion}

This work does not merely propose a more efficient routing algorithm; it instantiates a new governance architecture at the intersection of physics, law, and institutional decision-making. By integrating quantum-inspired collapse dynamics, embedded legal constraint operators, and trustworthy environmental feedback, the system realizes a \emph{law–technology co-design} paradigm underpinned by principles of \emph{multilevel governance} and shaped through \emph{mechanism design} to align incentives, ensure accountability, and adapt to evolving crises.

At the core is a layered coordination problem: multiple authorities (local, regional, federal), technical agents (sensing, decision engine, execution), and legal regimes must jointly produce timely, lawful, and robust pharmaceutical dispatches under uncertainty. The proposed framework addresses this by (1) embedding legal constraints as intrinsic operators rather than external vetoes, (2) using entropy-adjusted collapse timing to mediate commitment under informational asymmetry, and (3) employing a distributed provenance layer (blockchain) to reduce trust frictions across actors. These design choices create an incentive-compatible decision substrate: actors are guided toward socially desirable behaviors (compliance, transparency, prompt reporting) because deviations either degrade confidence (observable entropy) or surface in audit trails, converting latent risks into accountable signals.

\begin{figure}[ht]
  \centering
 \includegraphics[width=0.8\linewidth]{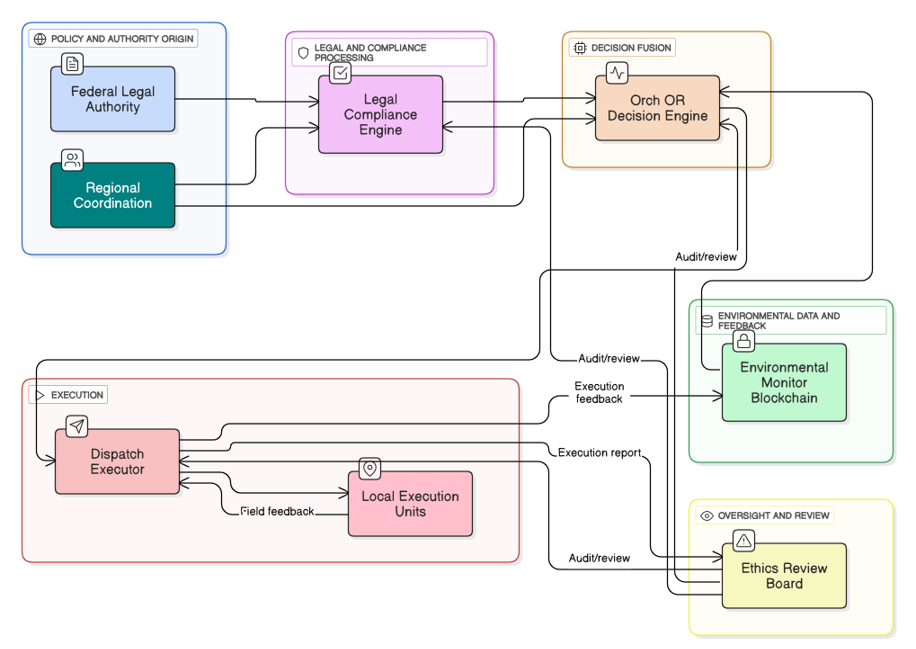}
  \caption{Responsibility and accountability in a multilevel governance structure: statutes and waivers originate at the federal/regional level; legal and environmental subsystems mediate constraints; the decision engine outputs routes; execution feeds back to oversight and sensing. Dashed arrows denote review/audit channels.}
  \label{fig:responsibility}
\end{figure}

\begin{figure}[ht]
  \centering
  \includegraphics[width=0.8\linewidth]{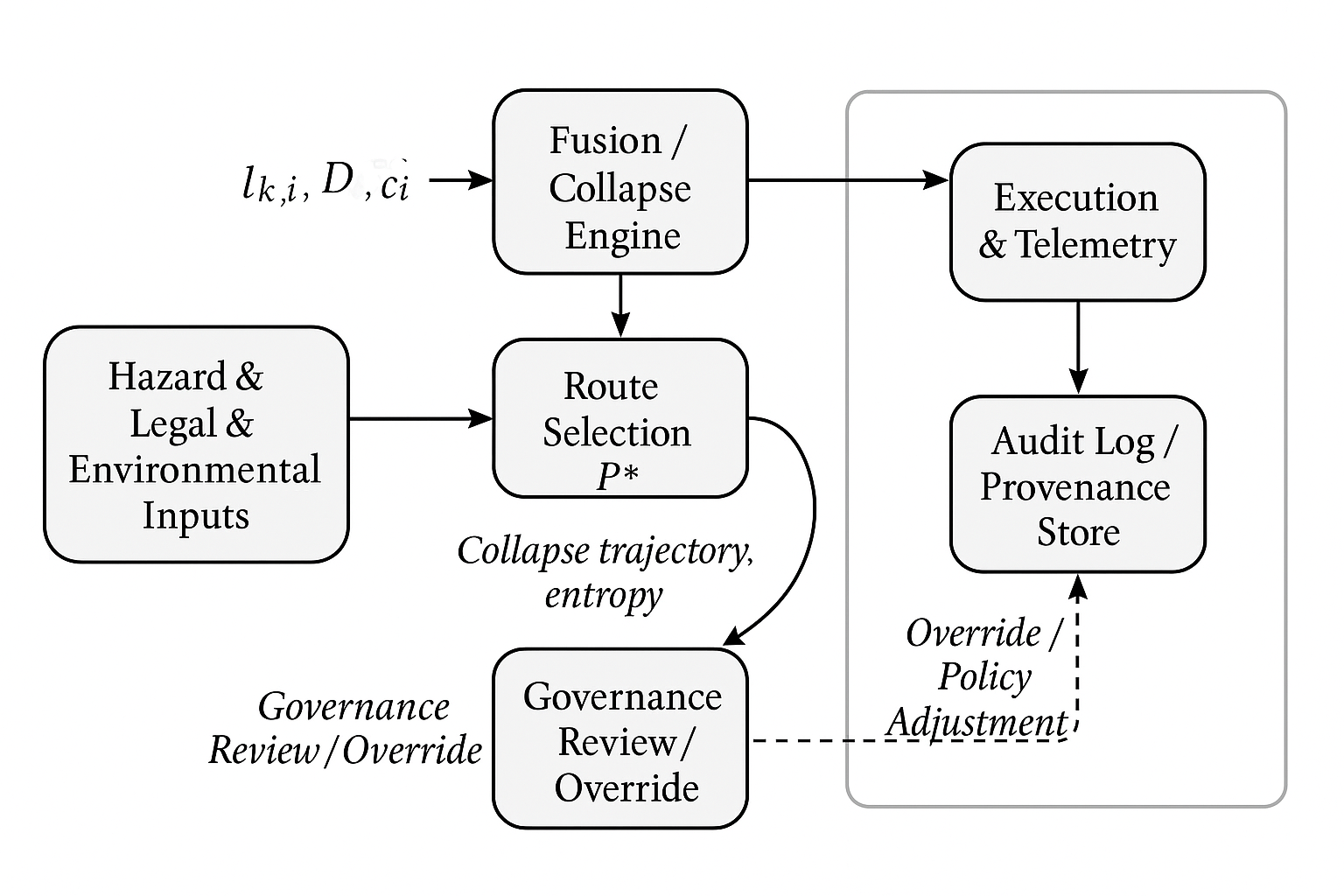}
  \caption{Decision and audit flow: inputs are fused, collapse dynamics produce a route; execution telemetry and internal signals (entropy, legal activations) are logged; governance actors can review and intervene, creating a feedback mechanism consistent with mechanism design for accountability.}
  \label{fig:auditflow}
\end{figure}

\begin{table}[ht]
\centering
\caption{Stakeholders, incentives, and governance metrics.}
\label{tab:stakeholders_metrics}
\begin{tabular}{p{3cm} p{3.5cm} p{3.5cm} p{3cm}}
\hline
Actor & Primary Incentive & Built-in Mechanism Alignment & Observable Metric \\
\hline
Federal Authority & Legal consistency, national reliability & Statute encoding, emergency waiver rules & Compliance rate, dispute frequency \\
Regional Coordinators & Rapid response, localized legitimacy & Utility weight tuning, conditional waivers & Response latency, override incidence \\
Legal Engine Operators & Correct interpretation & Projection operators with provenance & Rule activation traceability, correction rate \\
Technical Decision Unit & Optimal lawful dispatch & Entropy-aware collapse, feedback assimilation & Confidence entropy, selection stability \\
Environmental Trust Layer & Data integrity & Blockchain proofs, anomaly scoring & Provenance mismatch alerts, consensus health \\
Field Executors & Successful delivery & Dispatch commands with fallback options & Delivery success, deviation reports \\
Governance Review / Ethics & Equity and accountability & Review triggers on high entropy/overrides & Bias metrics, override justifications \\
\hline
\end{tabular}
\end{table}

The system’s design reflects classical \emph{mechanism design} desiderata: \emph{incentive compatibility} is achieved because deviation (e.g., suppressing provenance, misreporting legal context) reduces trust and can be detected through audit logs; \emph{participation constraints} are respected via transparent feedback that allows stakeholders to understand and contest decisions; and \emph{robustness} comes from layered fallbacks (e.g., conservative safe routes) and adaptive entropy thresholds that prevent rash commitment when signal coherence is low.

Policy-makers operate within a \emph{polycentric} governance environment where authority is distributed but interdependent. The collapse engine functions as a coordination mechanism that internalizes cross-level externalities: regional emergency waivers influence the feasible decision manifold, while local execution feedback updates priors used in future federal-level policy evaluation. This creates a dynamic subsidiarity loop in which decisions are made at the most local effective level, yet remain constrained by higher-order legal and normative structure.

Risk management in this architecture emphasizes converting latent ambiguity into explicit governance signals. High entropy becomes a \emph{trigger} for human review rather than epistemic paralysis; conflicting legal activations are surfaced as competing projection weights whose provenance and temporal context are stored for adjudication. The audit flow (Figure~\ref{fig:auditflow}) closes the accountability loop: every collapse trajectory, rule application, environmental input, and execution deviation is timestamped, enabling ex post causal tracing and forming the basis for adaptive regulation.

Implementation demands a careful calibration of \emph{policy knobs}—the weight vector \((\alpha,\beta,\delta)\), entropy sensitivity, and waiver thresholds—through \emph{scenario-based policy simulation} before live deployment to avoid perverse incentives (e.g., gaming legal exemptions to minimize latency). Equity considerations mandate periodic reweighting of historical priors \( c_i \) and institutional audits to detect systemic under-service, integrating fairness constraints into the optimization layer.

In summary, the proposed system constitutes a governance-aware infrastructure: it is not only a decision engine but a \emph{mechanism} that structures interactions among institutions, encodes law within computations, surfaces uncertainty for deliberation, and embeds auditability to sustain legitimacy over time.

\section{Conclusions and Future Work}
\label{sec:conclusion}

This work has developed and analyzed a novel governance-aware framework for emergency pharmaceutical dispatch that tightly fuses physical uncertainty, legal constraint embedding, and adaptive decision dynamics into a single operational substrate. By representing candidate routes as a superposed state and resolving them via entropy-modulated collapse, incorporating statutes as real-time operators that reshape feasible decision manifolds, and grounding environmental feedback in tamper-evident provenance, the system produces dispatch choices that are simultaneously responsive, legally compliant, robust to degradation, and explainable. The interplay among these components creates an incentive-compatible mechanism: ambiguity is surfaced (via entropy) to trigger oversight rather than failure, legal deviations are penalized yet flexibly handled through soft constraint modeling, and environmental risks attenuate unreliable paths while preserving traceability through distributed ledger auditing.

The primary theoretical contribution is a unified formalism in which delivery utility, legal fidelity, and environmental survivability co-determine selection drives in a nonlinear master equation whose collapse timescale adapts to information entropy. Algorithmically, this yields a decision engine that balances competing objectives without external arbitration, provides diagnostics (confidence, entropy trajectory, rule activation provenance), and supports human-in-the-loop governance through clearly defined intervention thresholds. Conceptually, the framework operationalizes law–technology co-design within a multilevel governance setting: higher-order statutes and waivers influence the collapse landscape, regional tuning parameters encode policy preferences, and local execution feedback updates priors and informs subsequent legal evaluation, forming a dynamic subsidiarity loop.

From a deployment perspective, the path forward begins with controlled pilots to calibrate the mapping between statute language and compliance operators, tune entropy sensitivity and waiver logic, and validate environmental sensing under real conditions. A policy sandbox should precede operational rollout, enabling stress-testing of utility weight configurations to surface and mitigate unintended incentives. Institutional integration requires embedding the decision substrate into existing emergency coordination chains, establishing audit repositories for rule provenance and collapse traces, and codifying override protocols tied to measurable uncertainty (e.g., sustained high entropy or conflicting legal signals). Scaling to broader and cross-jurisdictional contexts entails federated compliance layers that reconcile heterogeneous legal regimes and conflict-resolution hierarchies while preserving local adaptivity.

Key open challenges remain. Empirical grounding of the abstraction—especially the interpretation of quantum-style collapse in socio-technical settings with human actors and discretionary legal interpretation—requires longitudinal field data and adaptive refinement. Incorporating learning (e.g., constrained reinforcement learning or Bayesian updating) to evolve priors and utility weights while safeguarding fairness and legal fidelity is nontrivial and demands mechanisms to prevent drift or opacity. Formal verification of compliance guarantees, latency bounds under adversarial sensing corruption, and stability of collapse under contradictory inputs would strengthen trust and enable certified deployment. Hardware and software co-design opportunities exist to accelerate the decision dynamics, particularly for concurrent multi-incident scaling with tight latency requirements. Finally, generalizing the lawful collapse substrate to other high-stakes domains—such as epidemic resource allocation, infrastructure recovery, or climate-induced evacuation—offers a promising avenue for building a class of resilient, legitimate crisis governance systems.

In sum, by making law, risk, and decision-making co-constitutive rather than sequential, this framework establishes both a theoretical and practical foundation for emergency logistics that is measurable, auditable, and adaptable. The future work agenda—spanning empirical validation, policy tuning, federated legal harmonization, learning augmentation, and formal assurance—constitutes a roadmap for evolving this prototype into a deployable, multi-domain governance infrastructure.

\end{document}